# A NOVEL VARIABLE SELECTION METHOD BASED ON FREQUENT PATTERN TREE FOR REAL-TIME TRAFFIC ACCIDENT RISK PREDICTION

## Lei Lin[1], Qian Wang[1], and Adel W. Sadek[1]

[1] Department of Civil, Structural & Environmental Engineering,
Institute for Sustainable Transportation & Logistics
University at Buffalo, the State University of New York
233 Ketter Hall
Buffalo, NY 14260
e-mail: llin22@buffalo.edu
qw6@buffalo.edu
asadek@buffalo.edu

**Abstract.** *Traffic accident data are usually noisy, contain missing values, and heterogeneous. How to select the most important variables to improve real-time traffic accident risk prediction has become a concern of many recent studies. This paper proposes a novel variable selection method based on the Frequent Pattern tree (FP tree) algorithm. First, all the frequent patterns in the traffic accident dataset are discovered. Then for each frequent pattern, a new criterion, called the Relative Object Purity Ratio (ROPR) which we proposed, is calculated. This ROPR is added to the importance score of the variables that differentiates one frequent pattern from the others. To test the proposed method, a dataset was compiled from the traffic accidents records detected by only one detector on interstate highway I-64 in Virginia in 2005. This data set was then linked to other variables such as real-time traffic information and weather conditions. Both the proposed method based on the FP tree algorithm, as well as the widely utilized, random forest method, were then used to identify the important variables or the Virginia data set. The results indicate that there are some differences between the variables deemed important by the FP tree and those selected by the random forest method. Following this, two baseline models (i.e. a nearest neighbor (k-NN) method and a Bayesian network) were developed to predict accident risk based on the variables identified by both the FP tree method and the random forest method. The results show that the models based on the variable selection using the FP tree performed better than those based on the random forest method for several versions of the k-NN and Bayesian network models. The best results were derived from a Bayesian network model using variables from FP tree. That model could predict 61.11% of accidents accurately, while having a false alarm rate of 38.16%.*

## 1   INTRODUCTION

The cost of traffic accidents, in terms of loss of lives and property, is still unfortunately very high. According to the accident report of the United States Census Bureau, in 2009 alone, 10.8 million accidents occurred and about 35,900 persons were killed [1]. Motivated by the urgent need to reduce or eliminate such a cost, transportation researchers have for years been actively engaged in studies related to hazardous location/hot spot identification [2], accident injury-severities analysis [3], accident duration analysis [4] and so on. More recently, there has been an increased interest in developing real-time traffic accident risk prediction models which can provide estimates of the probabilities of traffic accident occurrence based on real-time traffic condition and weather data. The premise here is that once "high-risk" traffic patterns are identified, countermeasures may be activated in real-time to reduce that risk.

This paper proposes a novel variable selection method, based on the frequent pattern tree (FP tree) algorithm, which is used to identify the important variables for real-time traffic accident risk prediction models. Specifically, a new algorithm was designed, which can provide the variable importance score. We then compared the variable importance ranking results of our model with the random forest method, which is among the commonly used methods in traffic safety research to identify the relevant variables. Based on the set of variables identified as important by both our proposed method and the random forest method, we then developed and compared the performances of two base line models: a k nearest neighbor model and a Bayesian network, using a dataset extracted for a specific detector location on I-64 in Virginia. The results show that the model trained with the selected variables based on frequent pattern tree always performed the best.

The organization of this paper is as follows. First, in methodology section, we introduce the proposed variable importance calculation algorithm based on the FP tree algorithm. We then describe the traffic accident dataset used in this paper. In the model development section, the variable importance ranking results of the FP tree algorithm and random forest method are presented, and the traffic accident risk prediction results of the k nearest neighbor and Bayesian network models are compared. Finally, the study's conclusions and future work suggestions are summarized.

## 2   MODEL METHODOLOGY

### 2.1 Frequent-Pattern Tree

The FP tree algorithm was proposed by Han et al. [5] to provide a compact representation of all relevant frequency information in a database. Suppose       a set of transactions in database DB, and each transaction $Tran$ is a set of items,              . A pattern     is called a frequent pattern when its support, which refers to the frequency at which     appeared in the        transactions, is equal to or greater than the minimum support.

$$\frac{supp\ X}{TN} \geq \sigma \qquad (1)$$

where $\sigma$ is a threshold value defined by user.

After a FP tree is built, each branch of the tree will be a frequent pattern in this dataset.

### 2.2 Variable Importance Calculation

The following steps can be conducted to calculate the variable importance:

1. For each frequent pattern, calculate its *relative object purity ratio* (ROPR). ROPR refers to the absolute difference between the proportion of records in this frequent pattern which contains the object value of interest (in this paper the object value refers to the case when an accident happened) and the proportion of records containing the object value in the whole dataset.

2. For each frequent pattern, find out the exclusive nodes which differentiate this frequent pattern from the others. For each transaction *Tran* in DB, find its corresponding frequent pattern and exclusive nodes , for each item in *Tran*. If the item exits in , add the ROPR to the item's importance score, otherwise, keep it unchanged.

3. For each variable, calculate the importance score, which is the sum of the importance scores of its corresponding items. For example, the variable "traffic volume" may be clustered into three items: "volume low", "volume medium" and "volume high", then the importance score of "traffic volume" is the sum of the importance scores of these three items. In this paper, the transfer of continuous variables into discrete items is realized using the fuzzy c-means clustering method [6].

## 3   MODELING DATASET

The dataset used in this paper includes the traffic accident records in 2005 for a segment of interstate highway I-64 in Norfolk, Virginia. This dataset also contains the weather, visibility, traffic volume, speed and occupancy at a one minute resolution.

As a classification problem, the pre-crash condition and normal traffic condition have to be defined first. In this paper the pre-crash condition is defined as a 10 min time period starting 5 minutes before the accident. For the normal condition, we defined it as the same time period as the pre-crash condition, but taking place on the same day of the week from two weeks earlier to two weeks later than the day of week with accident. If there is an accident happened within one hour before or after the normal traffic condition data, these data would not be included [7].

It's worth noting that due to a large number of missing values in the data set considered, this study extracted the corresponding pre-crash and normal traffic data from only one detector, the one that reported the accident. The explanatory variables included eight variables: the mean of the weather type, the mean of visibility values, the mean and standard deviation of traffic volume, traffic speed, and occupancy in the defined time period. The response variable accident occurrence is a binary variable: 1 for the pre-crash data, and 0 for normal traffic data.

Overall, the accident dataset included 174 pre-crash records and 569 normal traffic records. 80% of pre-crash records and normal traffic records are randomly chosen to serve as the training dataset, and the remaining 20% are considered as the test dataset. Specifically, the training dataset included 139 pre-crash records and 455 normal traffic records; and the testing dataset included 35 pre-crash records and 114 normal traffic records.

## 4   MODEL DEVELOPMENT AND RESULTS

### 4.1 Variable Importance Calculation

The variable importance score was calculated following the proposed method described in section 2.2. For identifying the important variables using the Random Forest method, the package "randomForest" in the R software was used [8]. One importance criteria used in the "randomForest" method is the mean decrease of the Gini index, which measures the contribution of a variable to the homogeneity of the nodes and leaves in the random forest [9]. If one variable can result in nodes with higher purity, its impact on decreasing the Gini index will also be higher, which would mean that the variable is more important than others. The results of detecting the importance or significance of the variables are shown in Table 1 for both the FP-Tree algorithm and the Random Forest method.

| Variable | Importance | |
|---|---|---|
| | FP Tree | Random Forest |
| $Mean_{wea}$ | 37.6 | 9.88 |
| $Mean_{vis}$ | 30.8 | 13.39 |
| $Mean_{vol}$ | 48.6 | 27.51 |
| $Mean_{ocu}$ | 35.8 | 24.89 |
| $Mean_{spe}$ | 20.8 | 29.02 |
| $Std_{vol}$ | 42.8 | 29.15 |
| $Std_{ocu}$ | 15 | 26.41 |
| $Std_{spe}$ | 29 | 30.11 |

Table 1: Variable importance calculation results based on FP Tree and random forest.

As can be seen, there are differences between the results of the two methods. For example, according to the FP tree method, the mean speed ($Mean_{spe}$) and the standard deviation of the occupancy ($Std_{ocu}$) are the two variables with the least variable importance scores. The Random Forest method, on the other hand, indicates that the mean weather ($Mean_{wea}$) and mean visibility ($Mean_{vis}$) are the least important.

### 4.2 k Nearest Neighbor (k-NN)

Based on the important variables identified from both the FP-tree and Random Forest method, the study then developed a set of k-NN models for accident risk prediction. The results are shown in Table 2 for two k-NN models (with k set as 2 and 3) and for three different scenarios: (1) using all the variables available; (2) excluding the two least important variables identified by the FP-tree algorithm (i.e. excluding $Mean_{spe}$ and $Std_{ocu}$); and (3) excluding the two least important variables identified by the Random Forest method (i.e. excluding $Mean_{wea}$ and $Mean_{vis}$. Note that there are two criteria introduced in Table 2. Sensitivity means the probability of successful prediction when accidents happen. The false alarm rate refers to the probability of reporting an accident while there are no accidents in fact. This means that a good traffic accident risk prediction model should have a high sensitivity and a low false alarm rate.

| Variable selection | Criteria | k=2 | k=3 |
|---|---|---|---|
| All variables | Sensitivity | 40.00% | 48.57% |
| | False alarm rate | 42.98% | 56.14% |
| FP Tree | Sensitivity | 45.71% | 60.00% |
| | False alarm rate | 42.10% | 54.38% |
| Random Forest | Sensitivity | 37.14% | 54.28% |
| | False alarm rate | 49.12% | 61.40% |

Table 2: Performance of k-NN for different variable selection.

As can be clearly seen from Table 2, all models that used the FP-Tree variables outperformed those that used all the variables, and were based on the Random Forest results. However, it should be noted that the k-NN models do not appear to perform very well for this task. For example, when k was set to 2, the success rate was less than 46% for all the scenarios while the false alarm rate can be as high as 49.12%. The comparison between the k-NN with k=3 and the one with k=2 shows that adding one nearest neighbor in the method will significantly increase the prediction sensitivity. The downside is that the false alarm rates increase as well with the increase of the number of nearest neighbors of consideration.

### 4.3 Bayesian Network

The second model considered in this study was a Bayesian network, which requires the discretization of continuous variables as a first step. This step heavily relies on he researchers' judgments and objectives, and could have a major impact on model performance [7]. To realize the step, this paper applied the normalized equal distances (NED) method implemented in the software package, Bayesialab [10].

Different variations of the Bayesian network model were considered depending upon the equal width number (specifically the number was set to 3 and 4 separately). For the Bayesian network structure, here we consider one of the most plausible structures which let the response variable be the child node of the possible explanatory variables [7]. We also consider the three scenarios considered in relation to the k-NN models, (1) using all the variables available; (2) excluding the two least important variables identified by the FP-tree algorithm; and (3) excluding the two least important variables identified by the Random Forest method The software Netica was used for learning the Bayesian network parameters [11]. The Bayesian network using variables based on the FP tree and having the equal width number set to 4 is shown in Figure 1 as an example:

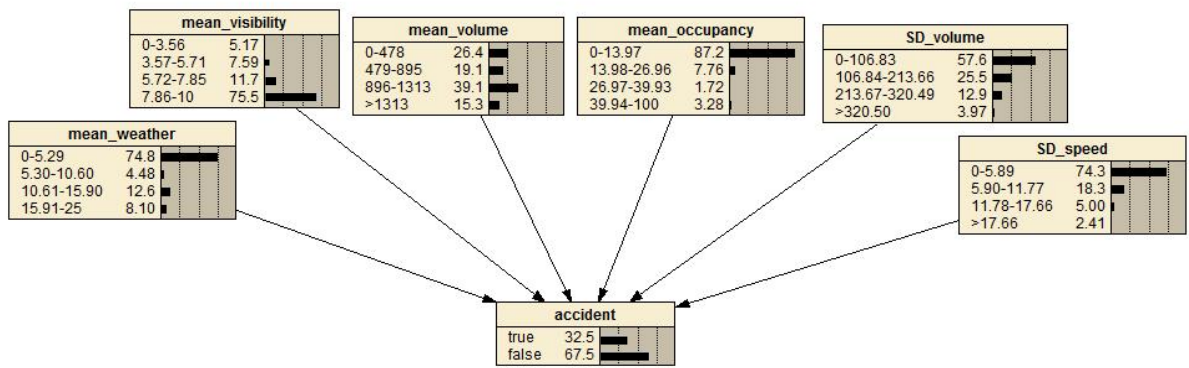

Figure 1: Bayesian network using variables based on FP tree.

As can be seen in Figure 1, each of the six explanatory variables is split into four intervals based on NED. The number behind each interval is the probability that the value of variable falls in that interval. For example, for the "mean_volume" node, we can see that 15.3% of records in the training dataset have the Mean$_{vol}$ greater than 1313 vehicles/hour.

The probability threshold of the Bayesian network is set to 0.2, which means if the probability of being true for the node "accident" is greater than 0.2, we assume that the output of the observation is that the accident would happen. The performance of the Bayesian networks for the testing dataset is shown in Table 3.

| Variable selection | Criteria | NED (3) | NED (4) |
|---|---|---|---|
| All variables | Sensitivity | 50.00% | 61.11% |
| | False alarm rate | 47.37% | 47.37% |
| FP Tree | Sensitivity | 44.44% | 61.11% |
| | False alarm rate | 38.15% | 38.16% |
| Random Forest | Sensitivity | 33.33% | 61.11% |
| | False alarm rate | 42.10% | 53.95% |

Table 3: Performance of Bayesian network for different variable selection.

Several findings can be deducted from Table 3. First, the number of NED can significantly affect the performance of the Bayesian network. The sensitivity of the model experiences an obvious improvement while the false alarm rate remains almost the same when the NED number is changed from 3 to 4 (the only exception is the case when the variables are selected based on the random forest method, where the false alarm rate also increased). Second, in terms of the variable selection method, the results confirms the observation made in relation to the k-NN models in that the Bayesian networks using the variables based on the FP tree performed better than the Bayesian networks with the variables identified based on the random forest based method. The best result of was obtained with a Bayesian network having NED equal to 4 and with variables selected based on the FP-Tree algorithm. For that network, the sensitivity was 61.11% and the false alarm rate was as low as 38.16%.

## 5 CONCLUSIONS

This study proposed a novel variable selection algorithm based on the FP tree algorithm for real-time traffic accident risk prediction modeling, and developed both k-NN models as well as Bayesian networks for real-time accident prediction using data from I-64 in Virginia. The performance of the FP-Tree method was also compared to that of the Random Forest method. Among the main conclusions of the study are:

1. The best accident prediction model was a Bayesian network model with NED number equal to 4 and using variables selected by the FP tree algorithm. The sensitivity of that model in terms of accurately predicting accident occurrence was 61.11% and the false alarm rate was 38.16%. Considering that data from only one detector were used in this study, these results are promising.

2. The performances of the k-NN models can be significantly affected by the selection of the number of neighbors, k. For the Bayesian network models, the discretization of variables also has an obvious impact on performance.

3. Regardless of the model used (i.e. whether k-NN or Bayesian network), the models based on the variables identified by the FP tree algorithm performed better than those based on the results from the random forest method.